\documentclass[draftcls, onecolumn, 12 pt]{IEEEtran}
\usepackage{amssymb, amsmath, graphicx, paralist,subfigure,cite,graphicx}
\usepackage{algorithm2e}
\usepackage{color}

\def\mb{\mathbf}

\def\mc{\mathcal}

\begin{document}
\title{Sensor Fault Detection and Isolation via Networked Estimation: Full-Rank Dynamical Systems}

\author{Mohammadreza Doostmohammadian,  Nader Meskin,   \textit{Senior Member, IEEE}
\thanks{This paper was supported  by the Qatar National Research Fund (a member of the Qatar Foundation) under the NPRP Grant number NPRP10-0105-170107.}
\thanks{Mohammadreza Doostmohammadian is with the Faculty of Mechanical Engineering at Semnan University, Semnan, Iran, email: \texttt{doost@semnan.ac.ir}. Nader Meskin is with the Electrical Engineering Department at Qatar University, Doha, Qatar, email: \texttt{nader.meskin@qu.edu.qa}.
}}
\maketitle

\begin{abstract}
	This paper considers the problem of simultaneous sensor fault detection, isolation,  and networked estimation of linear full-rank dynamical systems. The proposed networked estimation is a variant of single time-scale protocol and is based on (i) consensus on \textit{a-priori}  estimates and (ii) measurement innovation. The necessary connectivity condition on the sensor network and stabilizing block-diagonal  gain matrix is derived based on our previous works. Considering additive faults in the presence of system and measurement noise, the estimation error at sensors is derived and proper residuals are defined for fault detection. Unlike many works in the literature, no simplifying  upper-bound condition on the noise is considered and we assume  Gaussian system/measurement noise. A probabilistic threshold is then defined for fault detection based on the estimation error covariance norm. Finally, a graph-theoretic sensor replacement scenario is proposed to recover possible loss of networked observability due to removing the faulty sensor. We examine  the proposed fault detection and isolation scheme on an illustrative academic example to verify the results and make a comparison study with related literature.
	
	\textit{Keywords:} Fault Detection and Isolation, Observability, Networked Estimation, System Digraph, Residual
\end{abstract}

\section{Introduction}\label{sec_intro}
\IEEEPARstart{F}{ault} Detection and Isolation (FDI) is an emerging topic of recent literature in control and signal-processing \cite{hwang2009survey}. In general, a fault is characterized as an unpermitted deviation or malfunction of (at least) one of the system parameters/properties from its standard condition, which may take place at the plant, sensor, or actuator. Fault Detection, Isolation, and Reconfiguration (FDIR) is therefore a methodology to ensure acceptable system operation and to compensate for occurrence of faults detected by  FDI module\cite{hwang2009survey}. The fault diagnosis schemes are spanned from the centralized approaches \cite{davoodi2014simultaneous,jee2012h,li2009observer} to more recent distributed methods \cite{ferrari2011distributed,davoodi2013distributed,teixeira2014distributed,shames2011distributed,quan2016observer,marino2017distributed,li2016distributed}. Indeed, with recent technological trends, many practical systems of interest are either large-scale or physically distributed, and thus it is required to develop distributed or networked FDI strategies. Toward this goal, in this paper, the problem of simultaneous sensor fault detection, isolation and networked estimation of dynamical system is investigated.

 Fault detection and isolation in networked dynamical systems  has been extensively studied in the literature.
In \cite{davoodi2013distributed}, an FDI strategy is proposed for a distributed heterogeneous multi-agent system and
in \cite{teixeira2014distributed}, fault detection of a group of interconnected noise-free individual subsystems  is considered such that each subsystem  is able to detect faults in its neighborhood.  In \cite{shames2011distributed}, an FDI problem for interconnected double-integrator  system  under unknown system faults and fault-free measurements is investigated such that almost all faults except the faults with zero dynamics can be detected. Similarly, in \cite{quan2016observer}, a faulty second-order dynamical system (representing a heterogeneous multi-agent system) is considered with bounded disturbance and fault-free measurements and  a distributed observer-based FDI strategy  is proposed. The case of  homogeneous multi-agent linear systems under upper-bounded system noise is considered in \cite{marino2017distributed}, where  a distributed observer-based FDI  based on fault-free measurements  is developed to detect faults at agents. Development of FDI on noise-free faulty systems monitored by a multi-agent network communicating under an event-triggered framework is considered in \cite{li2016distributed}.
It should be noted that most of these works are with the main aim of fault detection and  the state estimation of the underlying dynamical system is not considered.

The other related trend in the literature is attack detection and resilient/secure estimation, both centralized \cite{chong2015observability,lee2015secure,ren2020secure,pajic2016attack,shoukry2017secure,pajic2014robustness,chen2016dynamic} and distributed \cite{deghat2019detection,guan2017distributed,su2019finite,an2019distributed,he2020secure,ao2018distributed,mitra2018secure,alanwar2019distributed,chen2018resilient2}. Distributed estimation strategies are more prominent in recent years due to emergence of large-scale applications. In \cite{deghat2019detection},  a distributed $H_\infty$ observer is proposed to detect \textit{admissible} biasing attacks over distributed estimation networks by introducing an auxiliary input tracking model. In \cite{guan2017distributed}, both false-data injection attack at the system/process and jamming attack at the communication network of sensors are considered and  in \cite{su2019finite}, a distributed gradient descent protocol is adopted to optimize the norm of the covariance of the measurement updates as the cost function.   In \cite{an2019distributed},  a distributed method is proposed to estimate the system state over a \textit{k-regular} sensor network under attacked noise-free measurements and in \cite{he2020secure}, a distributed resilient estimation scheme is proposed where the time-scale of the sensors (as estimators) needs to be faster than  the system dynamics. 

Secure distributed estimation of cyber-physical systems composed of interconnected subsystems each monitored by a sensor under possible unidentifiable attack is discussed in \cite{ao2018distributed}. In \cite{mitra2018secure}, secure distributed estimators for noise-free system and measurements is proposed over dynamic sensor networks subject to communication loss.
Other relevant works include secure estimation based on reachability analysis in \cite{alanwar2019distributed} and static parameter estimation in \cite{chen2018resilient2}. Note that most of these works propose a resilient estimation protocol under sensor attack, where no attack detection and isolation scheme is considered.
In general, distributed fault/attack detection finds application in cases where the centralized architecture is not possible or not desirable, as in smart-grid monitoring \cite{manandhar2014detection}, large-scale wind-farms \cite{wei2008fault}, and  networked unmanned vehicles \cite{meskin2011fault}.

In this paper,  a quantitative model-based method is adopted, as in the observer-based methods, to develop explicit mathematical model and
control theory to generate residuals for sensor fault detection. The proposed observer/estimator in this paper is distributed/networked, which finds application in large-scale architectures. Similar networked estimators are proposed in the literature \cite{garin2010survey,azizi2014networked,sauter:09,kar2012distributed,sayedtu12,battistelli_cdc,nuno-suff.ness}. A single time-scale networked estimator is developed to track the global state of the dynamical system over a distributed network of sensors each taking local measurements with partial observability. Unlike some  literature  \cite{sauter:09,kar2012distributed,azizi2014networked}, it is not assumed that the system is observable in the neighborhood of each sensor, i.e. the system is not necessarily observable by the measurements directly shared with the sensor. This significantly reduces the connectivity requirements on the sensor network, and therefore reduces the network communication costs.
Similar to \cite{sayedtu12,battistelli_cdc,nuno-suff.ness}, it is assumed that the underlying dynamical system is full-rank and conditions for networked observability are developed.

 The design procedure in this paper is based on structured system theory, previously developed in \cite{jstsp,isj_cyber}. Structural (or generic) analysis is widely used in system theory, namely in fault detection \cite{commault2008structural,commault2007sensor} and in distributed observability characterization \cite{globalsip14}. Using structural analysis, the structure of the sensor network and local estimator gain can be designed  to satisfy distributed observability. However, in this paper possible additive faults in sensors are considered and by mathematically deriving the estimation error and sensor residuals,  probabilistic thresholds on the residuals are obtained based on error covariance, which leads to the fault detection and isolation logic. Finally, after detecting  and isolating the faulty sensor,  a graph-theoretic method is proposed to replace the faulty sensor with an observationally equivalent state measurement to compensate the loss of observability. To summarize, the main contributions of this paper as compared with related literature are as follows:
\begin{itemize}	
	\item Considering system/process and measurement noise is a challenge in FDI strategies. This is mainly due to the fact that it is generally hard to distinguish between the presence of faults and system/measurement noise as both fault and noise terms may affect the sensor residuals. In this direction, some works in the literature \cite{teixeira2014distributed,chen2016dynamic,li2016distributed,mitra2018secure} assume that the system and/or sensor measurement are  noise-free, which is a simplifying assumption. In this paper, we consider both system and measurement noise and propose a probability-based threshold design on the residuals to overcome this challenge.
	
	\item Following the above  comment, many works in FDI and attack detection literature \cite{pajic2015attack,chong2015observability,lee2015secure,kodakkadan2017observer,Riccati-weakcons,marino2017distributed} assume that the noise variable is upper-bounded, i.e the noise term instead of taking different values, for example, from Gaussian distribution, only takes values in a limited range. This simplifying assumption helps to design deterministic thresholds on the residuals. In this paper, no such assumption is made, and the noise is assumed to be Gaussian random variable with no upper-bound, which is more realistic as compared to the mentioned references.
	
	\item In this paper,  a general LTI system is considered, which generalizes the multi-agent systems or interconnected systems each possessing a separate dynamics considered in \cite{davoodi2013distributed,teixeira2014distributed,marino2017distributed},  and the double-integrator system in \cite{shames2011distributed,quan2016observer}. Furthermore, the sensor network is considered as a Strongly-Connected (SC) graph as compared to restrictive regularity condition considered in \cite{an2019distributed}. 

	\item We adopt a graph-theoretic approach to recover the loss of observability in case of detecting and removing a faulty sensor. This approach is based on our previous works on observational equivalence \cite{icassp2016,spl17}.
\end{itemize}

The rest of this paper is organized as follows. In Section~\ref{sec_frame},  the general framework and state the problem are presented. Section~\ref{sec_est} provides the networked estimator scenario and develops the condition for error stability. In Section~\ref{sec_gain}, the algorithm for block-diagonal gain design at sensors is provided. Section~\ref{sec_fault} develops the sensor fault-detection and isolation logic, and in Section~\ref{sec_compens}, a graph-theoretic method for observability compensation is discussed. Section~\ref{sec_sim} gives simulation example to illustrate  the results. Finally, Section~\ref{sec_con} concludes the paper.

\section{The Framework} \label{sec_frame}
We consider noisy discrete-time linear systems as,
\begin{eqnarray}\label{eq_sys1}
\mb{x}_{k+1} = A\mb{x}_k + \mb{\nu}_k,\qquad k\geq0,
\end{eqnarray}
where $\mb{x}_k \in \mathbb{R}^n$ is the system state  and $ \mb{\nu}_k = \mc{N}(0,Q)$ is the system noise at time-step $k$. In this paper,  the underlying system matrix $A$ is considered to be full-rank and examples of such full-rank systems are \textit{self-damped dynamical systems} \cite{tnse19}. It should be noted that   the LTI system~\eqref{eq_sys1} can be also obtained by the  discretization of a continuous-time LTI system, based on  Euler  or Tustin  discretization methods discussed in \cite{tnse19} and both of these methods result in a full-rank discrete-time LTI system. Moreover,  the full-rank condition can be inherent to the system dynamics, such as in Nearly-Constant-Velocity (NCV) model for target dynamics in distributed tracking scenarios \cite{ennasr2018distributed}. In \cite{ennasr2018distributed}, the target is modeled as a discrete-time dynamical system whose associated matrix is full-rank due to its non-zero diagonal entries. The system full-rank condition is also considered in the networked estimation literature as in \cite{sayedtu12,battistelli_cdc,nuno-suff.ness}.

 The noise/fault-corrupted measurements of the system are taken by $N$ sensors,
\begin{align}\nonumber
\left(
\begin{array}{c}
\mb{y}^1_{k}\\
\vdots\\
\mb{y}^N_{k}
\end{array}
\right) &=
\left(
\begin{array}{c}
C_{1}\\
\vdots\\
C_{N}
\end{array}
\right)\left(
\begin{array}{c}
\mb{x}^1_{k}\\
\vdots\\
\mb{x}^n_{k}
\end{array}
\right)+
\left(
\begin{array}{c}
\mb{\zeta}^1_{k}\\
\vdots\\
\mb{\zeta}^N_{k}
\end{array}
\right) + \left(
\begin{array}{c}
\mb{f}^1_{k}\\
\vdots\\
\mb{f}^N_{k}
\end{array}
\right)
\end{align}
which can be written as the global measurement equation as follows:
\begin{eqnarray}
\mb{y}_k &=& C\mb{x}_k + \mb{\zeta}_k+\mb{f}_k, \label{eq_C}
\end{eqnarray}
where $\mb{y}_k \in \mathbb{R}^N$, $ \mb{\zeta}_k = \mc{N}(0,R)$ is the measurement noise, and $\mb{f}_k$ represents the sensor fault vector.
It is assumed that $\mb{\zeta}_k$ and $\mb{\nu}_k$ are zero-mean Gaussian while noise, i.e. $\mathbb{E}(\nu_k) = 0$, $\mathbb{E}(\zeta^i_k) = 0$, for all $i, k$, and  $\mathbb{E}(\nu_k\nu_m) = 0$, for all $k \neq m$, and similarly, for the measurement noise. Further, without loss of generality, it is assumed that each sensor measures one  of the system states.

The general problem in this paper is to design a stable networked estimation protocol in the absence of faults, and, then, a fault detection and isolation logic such that to detect possible faults in sensors. Given the system and measurements as in \eqref{eq_sys1} and \eqref{eq_C},  a group of sensors is considered, each embedded with communication and computation equipment  to measure a state of the dynamical system and process the measured data and information received from the other sensors to estimate the global state of the dynamical system.

Note that unlike many works in the literature \cite{sauter:09,kar2012distributed,azizi2014networked},  no assumption on the observability of system in the direct neighborhood of each sensor is considered which implies that the minimal connectivity on the communication network of agents is required. Each sensor adopts a single time-scale networked estimation protocol and by proper design of communication network structure and feedback gain matrix, each sensor tracks the system state by bounded steady-state estimation error in the fault-free case. In case of fault occurrence, i.e. $\mb{f}_k^i \neq 0$ at any sensor $i$, a residual-based fault detection  and isolation logic is proposed to detect the faulty sensor by comparing the sensor residuals with pre-specified thresholds. Note that
unlike many work in the literature \cite{pajic2015attack,chong2015observability,lee2015secure,kodakkadan2017observer,Riccati-weakcons}, it is not assumed that the system noise and/or measurement noise are upper-bounded. Finally, by detecting the faulty sensor, one may compensate for possible loss of observability in the distributed system by introducing a new sensor measurement replacing the faulty sensor.

\section{Networked Estimation Protocol} \label{sec_est}
In this section, we present our main tool to perform diagnosis as a single time-scale networked estimation protocol and analyze its error stability criteria. The networked estimator is proposed based on collaborative consensus on the \textit{a-priori}  estimates at sensors. The estimator is based on two steps as follows,
\begin{align}\label{eq_p}
\widehat{\mb{x}}^i_{k|k-1} =& \sum_{j\in\mathcal{N}_\beta(i)} W_{ij}A\widehat{\mb{x}}^j_{k-1|k-1},
\\\label{eq_m}
\widehat{\mb{x}}^i_{k|k} =& \widehat{\mb{x}}^i_{k|k-1} + K^i C_i^\top \left(\mb{y}^i_k-C_i\widehat{\mb{x}}^i_{k|k-1}\right),
\end{align}
where $\widehat{\mb{x}}^i_{k|k-1}$ is the estimate of state $\mb{x}_k$ at sensor $i$ given all the measurements up to time $k-1$, $\widehat{\mb{x}}^i_{k|k}$ is the estimated state given all the measurements up to time $k$, the matrix $W$ is a stochastic matrix for consensus on \textit{a-priori}  estimates, and $K^i$ is the gain matrix at sensor $i$. Note that the row-stochastic condition on $W$ is a necessary condition for (consensus) averaging of \textit{a-priori}  estimates. Recall that a matrix is row-stochastic if the summation of the entries in each row are equal to $1$, i.e. $\sum_{j=1}^{N} W_{ij}=1$. Further, $\mathcal{N}_\beta(i)$ defines the neighborhood of sensor $i$ over which the sensor shares \textit{a-priori}  estimates with the neighboring sensors and this neighborhood follows the structure of matrix $W$. In fact $\mathcal{N}_\beta(i)=\{j|j\rightarrow i\} \cup \{i\}$ where $j\rightarrow i$ implies that sensor $j$ sends its information to $i$ and  $\mathcal{N}_\beta(i)$ also includes the sensor $i$ itself, i.e. all the diagonal entries of the matrix $W$ are non-zero. This is due to the fact that each sensor uses its own estimate at previous time $\widehat{\mb{x}}^i_{k-1|k-1}$ to calculate its \textit{a-priori}  estimate $\widehat{\mb{x}}^i_{k|k-1}$. It should be noted that the protocol \eqref{eq_p}-\eqref{eq_m} differs from our previous works \cite{jstsp,isj_cyber} in two aspects: (i) the protocol in \cite{jstsp,isj_cyber} has one more consensus step on the measurement fusion in which the neighboring measurements are shared over a different (hub-based) network, while the protocol \eqref{eq_p}-\eqref{eq_m} only includes one step of averaging on \textit{a-priori} estimates. Therefore, the network connectivity condition in this paper is more relaxed as compared to  \cite{jstsp,isj_cyber}; (ii) the distributed approach in \cite{jstsp,isj_cyber} is fault-free where the sensor measurements are not accompanied with additive faults and consequently, the error dynamics analysis is different from this work.

\begin{figure}
	\centering
	\includegraphics[width=3in]{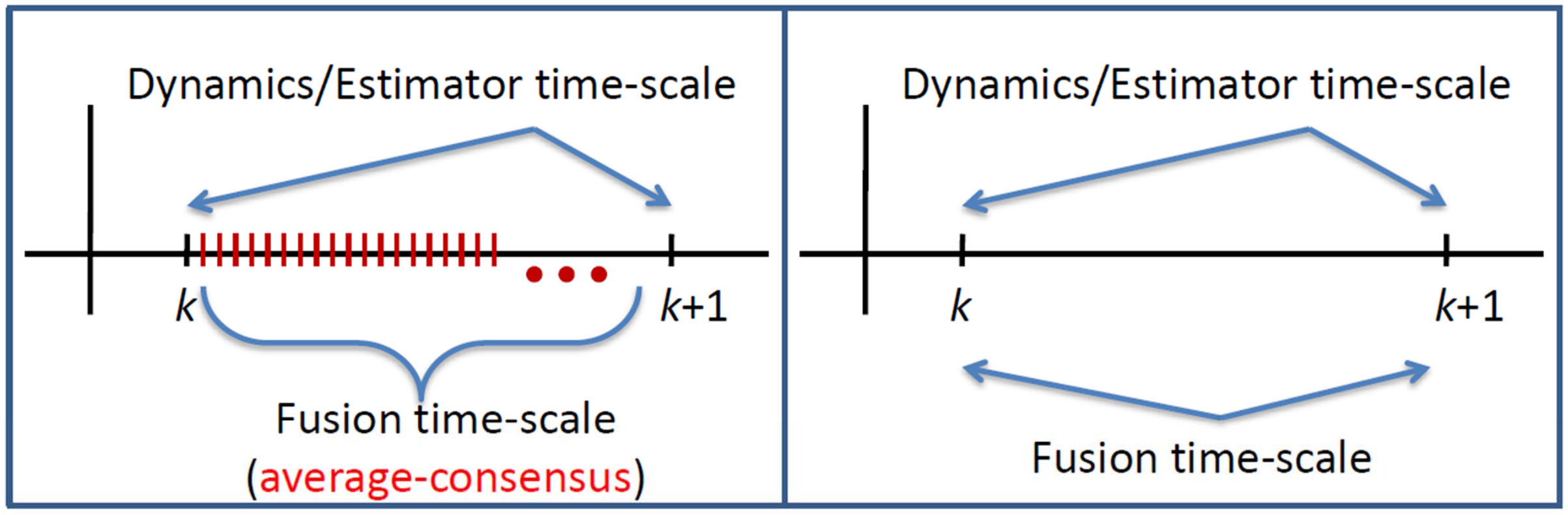}
	\caption{Single and multiple time-scale strategies for networked estimation: (Left) multi time-scale  also known as average-consensus based approach, and (Right) single time-scale approach. As it is clear from the figure, in the former scenario many steps of averaging is performed between every two successive time-steps of system dynamics, while in the latter case only one step of averaging is done between every two successive steps of system dynamics. The former requires faster processing units, while the latter is more desirable in real-time applications.}
	\label{fig_fusion}
\end{figure}

Note that the estimator \eqref{eq_p}-\eqref{eq_m} is single time-scale as compared to multi time-scale networked estimation given in \cite{olfati:05,he2020secure} where many step of averaging (or consensus) is performed between every two steps $k$ and $k+1$ of system dynamics (see Fig.~\ref{fig_fusion}).
It is known that the single time-scale approach is privileged over the multi time-scale method, since  the latter requires a  large number of information-exchange and communication over the sensor network between every two successive time-steps of the system dynamics. This implies that the communication time-scale needs to be much faster than the dynamics, which is not desirable and even may not be feasible for many large-scale real-time systems. To elaborate this, note that in protocol \eqref{eq_p}-\eqref{eq_m} only one step of averaging on \textit{a-priori}  estimates is done between two successive steps $k$ and $k+1$, while the protocol in \cite{olfati:05} requires many steps of consensus  between steps $k$ and $k+1$. In the multi time-scale estimation, the connectivity of the sensor network is more relaxed due to more information exchanges among the sensors, while in the single time-scale method, the sensor network requires more connectivity particularly for rank-deficient dynamical systems. In fact, it is proved that a SC network may not guarantee error stability for rank-deficient dynamical systems and certain hub-based network design is necessary, see \cite{jstsp,isj_cyber} for details.

Define the error $\mb{e}_{k}^i$ at time-step $k$ at  sensor $i$ as,
\begin{align}\label{love}
\mb{e}_{k}^i =& \mb{x}_{k|k} - \widehat{\mb{x}}^i_{k|k}, \nonumber
\\ \nonumber
  =&\mb{x}_{k} - \Bigl(\widehat{\mb{x}}^i_{k|k-1} + K^iC_i^\top (\mb{y}^i_k-C_i\widehat{\mb{x}}^i_{k|k-1})\Bigr)
\\\nonumber
  =&\mb{x}_{k} - \Bigl(\sum_{j\in\mathcal{N}_\beta(i)} W_{ij}A\widehat{\mb{x}}^j_{k-1|k-1} \nonumber \\ &+ K^i C_i^\top
\Bigl(\mb{y}^i_k-C_i\sum_{j\in\mc{N}_\beta (i)} W_{ij}A\widehat{\mb{x}}^j_{k-1|k-1}\Bigr)\Bigr).
\end{align}
By substituting~\eqref{eq_sys1}-\eqref{eq_C} in above, it follows that:
\begin{align}
\mb{e}_{k}^i =& A\mb{x}_{k-1} + \mb{\nu}_{k-1} - \Bigl(\sum_{j\in\mathcal{N}_\beta(i)} W_{ij}A\widehat{\mb{x}}^j_{k-1|k-1} \nonumber \\ &+ K^i C_i^\top
\Bigl(C_i\mb{x}_k + \mb{\zeta}^i_k+\mb{f}^i_k \nonumber \\ &- C_i\sum_{j\in\mc{N}_\beta (i)} W_{ij}A\widehat{\mb{x}}^j_{k-1|k-1}\Bigr)\Bigr) \nonumber
\\=&A\mb{x}_{k-1}+\mb{\nu}_{k-1} - \sum_{j\in\mathcal{N}_\beta(i)} W_{ij}A\widehat{\mb{x}}^j_{k-1|k-1} \nonumber \\
&-K^i \Bigl( C_i^\top C_i(Ax_{k-1}+\mb{\nu}_{k-1}) + C_i^\top \mb{\zeta}^i_{k} \nonumber \\
&+ C_i^\top \mb{f}^i_{k}-C_i^\top C_i\sum_{j\in \mathcal{N}_\beta(i)} W_{ij}A\widehat{\mb{x}}^j_{k-1|k-1}\Bigr) \nonumber
\end{align}
Recall that the row-stochastic condition of $W$ implies that $\sum_{j=1}^{N} W_{ij}=1$. Note that for $j\in \mathcal{N}_\beta(i)$, $W_{ij} \ne 0$ and for  $j\notin \mathcal{N}_\beta(i)$, $W_{ij} = 0$. Therefore, the row-stochastic condition can be re-written as $\sum_{j \in \mathcal{N}_\beta(i)} W_{ij}=1$. Based on this fact, it follows that (see \cite{isj_cyber} for similar analysis):
\begin{eqnarray}
A\mb{x}_{k-1}=\sum_{j\in \mathcal{N}_\beta(i)} W_{ij}A\mb{x}_{k-1}, \nonumber
\end{eqnarray}
and consequently:
\begin{align*}
\mb{e}_{k}^i &=\sum_{j\in \mathcal{N}_\beta(i)} W_{ij}A\mb{x}_{k-1} - \sum_{j\in\mathcal{N}_\beta(i)} W_{ij}A\widehat{\mb{x}}^j_{k-1|k-1} \nonumber \\
&-K^i C_i^\top C_i\Bigl(\sum_{j\in \mathcal{N}_\beta(i)}  W_{ij}A\mb{x}_{k-1} -\sum_{j\in \mathcal{N}_\beta(i)} W_{ij}A\widehat{\mb{x}}^j_{k-1|k-1} \Bigr) \nonumber \\&+ \mb{\nu}_{k-1}-K^iC_i^\top \mb{\zeta}^i_{k} -
K^i C_i^\top \mb{f}^i_{k} - K^i C_i^\top C_i\mb{\nu}_{k-1} \nonumber \\
&=\sum_{j\in \mathcal{N}_\beta(i)} W_{ij}A\mb{e}_{k-1}^j -K^i C_i^\top C_i\sum_{j\in \mathcal{N}_\beta(i)}  W_{ij}A\mb{e}^j_{k-1}+ \mb{\eta}_{k}^i ,
\end{align*}
where $\mb{\eta}_{k}^i$ collects the noise terms and fault term as follows:
\begin{eqnarray}
\mb{\eta}_{k}^i=\mb{\nu}_{k-1}-K^i \Bigl(C_i^\top \mb{\zeta}^i_{k} +
C_i^\top \mb{f}^i_{k}+ C_i^\top C_i\mb{\nu}_{k-1}\Bigr).
\end{eqnarray}
The collective error at the group of sensors is defined as the concatenation of errors at all sensors, i.e. $\mb{e}_{k}=(\mb{e}_{k}^{1\top},\dots,\mb{e}_{k}^{N\top})^\top$. For the collective error, it follows that:
\begin{eqnarray}\label{eq_err1}
\mb{e}_{k} = \widehat{A}\mb{e}_{k-1} +
\mb{\eta}_k,
\end{eqnarray}
where $\widehat{A}=W\otimes A - KD_C(W\otimes A)$ with $K=\mbox{blockdiag}(K^i)$ and $D_C=\mbox{blockdiag}(C_i^\top C_i)$. The collective noise vector is given as:
\begin{align}
\mb{\eta}_k &= \mathbf{1}_N \otimes \mb{\nu}_{k-1}- K D_C(\mathbf{1}_N \otimes \mb{\nu}_{k-1}) - K\overline{D}_C\mb{\zeta}_{k} -K\overline{D}_C\mb{f}_{k},
\label{eq_eta}
\end{align}
where $\mathbf{1}_N$ is the vector of 1s of size $N$ and $\overline{D}_C=\mbox{blockdiag}(C_i^\top)$.

To stabilize the error dynamics \eqref{eq_err1}, it is necessary that the pair $(W\otimes A, D_C)$ to be observable \cite{bay}. Note that the observability analysis of Kronecker product of matrices (and the associated composite graph) is discussed in details in \cite{kronecker_TSIPN}, where it is proved that for observability  of  $(W\otimes A, D_C)$, the matrix $W$ needs to be irreducible. This implies that the sensor network needs to be Strongly-Connected (SC). In fact,  this is also discussed in \cite{jstsp},  and since the protocol \eqref{eq_p}-\eqref{eq_m} is a variant of the protocol in \cite{jstsp}, the same analysis can be adopted here to prove the irreducibility of $W$ matrix. Therefore, any irreducible matrix (associated with a SC network) with non-zero diagonal entries while satisfying the row-stochastic condition can work as $W$ matrix. The SC connectivity of the sensor network is also considered in similar networked estimation literature, see for example \cite{sayedtu12,battistelli_cdc,nuno-suff.ness }.
Note that the $(W\otimes A, D_C)$-observability is also referred as networked observability (or distributed observability) condition. As it is explained in the next section, having $(W\otimes A, D_C)$-observability satisfied, using Linear-Matrix-Inequalities (LMI), the block-diagonal gain matrix $K$ can be designed such that $\widehat{A}$ is a Schur matrix \cite{jstsp,usman_cdc:11,rami:97}, i.e. $\rho(\widehat{A})<1$ where $\rho$ defines the spectral radius of a matrix.
\section{Design of Block-Diagonal Feedback Gain based on LMI Approach}\label{sec_gain}
In this section, we discuss the methodology for computation of a block-diagonal estimator gain~$K$. Note that, the observability of $(W \otimes A, D_C)$ in the distributed estimator \eqref{eq_p}-\eqref{eq_m} guarantees the existence of a full matrix~$K$, such that~$\rho (W \otimes A -K D_C (W\otimes A))<1$.
However, for having a distributed approach, the gain matrix is required to be block-diagonal. Such~$K$ is known to be the solution of an LMI as follows:
\begin{equation}\label{LMI-9}
\begin{aligned}
& ~~ \left( \begin{array}{cc} X&\widehat{A}^\top X\\ X\widehat{A}&X\\ \end{array} \right) \succ 0\Rightarrow \left( \begin{array}{cc} X&\widehat{A}^\top\\ \widehat{A}&Y\\ \end{array} \right) \succ 0,
\end{aligned}
\end{equation}
where~$\widehat{A} = W \otimes A -K D_C (W\otimes A)$ and ~$X \succ 0$ with $\succ$ denoting positive definiteness. Note that the left-hand-side of the above equation is nonlinear in~$K$; however its equivalent solution is proposed in the literature \cite{pang:95,5717159} as the right-hand-side in \eqref{LMI-9} with~$X=Y^{-1}$. Now, having the above LMI to be linear in~$K$, we note that the constraint~$X=Y^{-1}$ is a non-convex constraint. However, this constraint can be approximated with a linear function of
the matrices~$X,Y \succ 0$, satisfy~$X=Y^{-1}$ as the optimal point of the following optimization problem \cite{rami:97}:
\begin{equation}
\begin{aligned}
\displaystyle
\min
~~ &  \mathbf{trace}(XY) \\
\text{s.t.} ~~ & \left( \begin{array}{cc} X&I\\ I&Y\\ \end{array} \right) \succ 0,\\
~~ & K\mbox{~is~block-diagonal}.\\
\end{aligned}
\end{equation}
with~$X,Y \succ 0$. Overall, the optimization is summarized as follows; with~$(W \otimes A,D_C)$~observability, the  gain matrix $K$ is the solution to the following:
\begin{equation} \label{eq_min}
\begin{aligned}
\displaystyle
\min
~~ &  \mathbf{trace}(XY) \\
\text{s.t.}  ~~& X,Y\succ 0,\\ ~~ & \left( \begin{array}{cc} X&\widehat{A}^\top\\ \widehat{A}&Y\\ \end{array} \right) \succ 0,\\
~~& \left( \begin{array}{cc} X&I\\ I&Y\\ \end{array} \right) \succ 0,\\
~~ & K\mbox{~is~block-diagonal}.\\
\end{aligned}
\end{equation}
It should be mentioned that a solution of the second LMI is equivalent to~$X=Y^{-1}$, which results in an optimal value for minimum trace as~$nN$. Furthermore,~$\mathbf{trace}(XY)$ can be replaced with the linear approximation~$\mathbf{trace}(Y_0X+X_0Y) /2$ \cite{rami:97}, and the iterative Algorithm~\ref{algorithm} can be used to minimize this problem under given constraints.
\begin{algorithm} \label{algorithm}
	\textbf{Given:} $A,W,D_C$\\
	\KwResult{Gain matrix $K$}
	Find feasible points~$X_0,Y_0,K$\;
	\While{$\rho (\widehat{A}) >1$}{
		minimize~$\mathbf{trace}(Y_tX + X_tY)$ under the constraints given in equation~\eqref{eq_min} and find~$X,Y,K$\;
		$Y_{t+1}=Y$\;
		$X_{t+1}=X$\;
		$t=t+1$\;
	}
	\caption{Iterative calculation of block-diagonal gain~$K$.}
\end{algorithm}
\\
In \cite{rami:97}, it is shown that~$\mathbf{trace}(Y_tX + X_tY)$ is non-increasing and converges to~$2nN$. In this regard, a stopping criterion of the algorithm is established as  reaching within~$2nN + \epsilon$ of the
trace objective. Interested readers may refer to~\cite{usman_cdc:11,pang:95,5717159,rami:97, siljak08} for more details. It should be noted that this algorithm (and in general similar cone-complementarity algorithms) are of polynomial-order complexity, see \cite{ye1993fully,nesterov1994interior}.

\section{Sensor Fault Detection and Isolation}\label{sec_fault}
In this section, we present our results on the development of sensor fault  detection and isolation scheme for the considered system. Following the terminology in \cite{sundaram_lecture}, given the estimated state $\widehat{\mb{x}}_{k|k}^i$, define the estimated output at sensor $i$ as $\widehat{\mb{y}}_{k}^i = C_i \widehat{\mb{x}}_{k|k}^i$. Note that for fault-free case $\mb{y}_k^i-\widehat{\mb{y}}_{k}^i $ is steady-state stable and bounded by proper design of $W$ and $K$ matrices.
Thus, define the residual signal at sensor $i$ at time-step $k$ as,
\begin{align}\nonumber
r_k^i =& |\mb{y}_k^i-\widehat{\mb{y}}_{k}^i|=|\mb{y}_k^i-C_i \widehat{\mb{x}}_{k|k}^i|=
|C_i \mb{e}_k^i+\mb{\zeta}^i_k+\mb{f}^i_k|\\
=&|C_i \widehat{A}_i \mb{e}_{k-1}+C_i \mb{\eta}_k^i+\mb{\zeta}^i_k+\mb{f}^i_k|
\label{eq_rsidual}
\end{align}
where $\widehat{A}_i$ is the $i$th hyper-row  of the matrix $\widehat{A}$ defined as the block of rows from row $(i-1)n+1$ to row $in$. Moreover,  $C_i \mb{\eta}_k^i$ can be written  as:
\begin{align} \nonumber
C_i \mb{\eta}_k^i=&C_i \mb{\nu}_{k-1}-C_i K^i C_i^\top \mb{\zeta}^i_{k} -
C_i K^i C_i^\top \mb{f}^i_{k}\\ &-C_i K^i C_i^\top C_i\mb{\nu}_{k-1}.
\label{eq_eta_res}
\end{align}
Assume that sensor $i$ is faulty at some time-interval, i.e. $\mb{f}_k^i \neq 0$ for $k_1<k$. Note that, considering \eqref{eq_rsidual} and \eqref{eq_eta_res}, the estimation error is now affected by the sensor fault. In this case for the residual at sensor $i$, the term $C_i K^i C_i^\top \mb{f}^i_{k}\neq 0$  and $\mb{f}^i_{k}\neq 0$ in \eqref{eq_rsidual} and \eqref{eq_eta_res}, while for other sensors $C_j K^j C_j^\top \mb{f}^j_{k}=0$. Note that among many possible choices for the gain matrix $K$, it can be designed such that the term $C_i K^i C_i^\top$ is large enough to make $r_k^i$ more affected by the fault $\mb{f}^i_{k}$. Thus, although $\mb{f}^i_{k}$ may also appear in $\eta_{k-1}$ and $\mb{e}_{k-1}$, since $\rho(\widehat{A})<1$ the term $C_i\widehat{A}_i \mb{e}_{k-1}$ in \eqref{eq_rsidual} remains small and negligible as compared to larger values of $\mb{f}^i_{k}$ and $C_i K^i C_i^\top \mb{f}^i_{k}$. 
Therefore, by Schur stability of matrix $\widehat{A}$ and proper choice of the gain matrix $K$, the residual at faulty sensor $i$ is more affected while for the other sensors $j\neq i$, the  residuals are less affected since $\mb{f}^j_k=0$.  The
fault detection and isolation logic is therefore as follows: \textit{if the residual $r_k^i$ at sensor $i$ exceeds a predefined threshold then sensor $i$ is faulty.}
Thus, we need to define the thresholds on residuals for fault detection and isolation. In this direction, considering Gaussian noise $\mc{N}(0,Q)$ for system dynamics and $\mc{N}(0,R)$ for measurement noise, first the variance of the estimation error $\mb{e}_k^i$ is obtained and consequently  one can define a threshold such that to detect faults whenever the residual exceeds this threshold.

Let $P_k = \mathbb{E}(\mb{e}_k\mb{e}_k^{\top})$ and $\Sigma =\mathbb{E}(\eta_k\eta^\top_k) $. Then, it follows that:
\begin{align} \nonumber
   P_{k} =& \widehat{A}P_{k-1}\widehat{A}^\top + \Sigma \\
   =& \widehat{A}^kP_{0}(\widehat{A}^\top)^k + \sum_{j=0}^{k-1}  \widehat{A}^{j}\Sigma(\widehat{A}^\top)^{j}.
\end{align}
Recall that $\rho(\widehat{A})<1$, in the steady-state we have,
\begin{eqnarray} \nonumber
P_{\infty} = \lim_{k \rightarrow \infty} P_{k} =  \sum_{j=0}^{\infty}  \widehat{A}^{j}\Sigma(\widehat{A}^\top)^{j}.
\end{eqnarray}
Let $b = \|\widehat{A}\|_2<1 $, then using the result of \cite{khan2014collaborative} it can be proved that,
\begin{eqnarray} \label{eq_pinfty1}
\|P_{\infty}\|_2  \leq \frac{\|\Sigma\|_2}{1-b^2}.
\end{eqnarray}
On the other hand, for the fault-free case,
\begin{align} \nonumber
 \eta_k\eta^\top_k =& (I_{Nn}- K D_C)(\mathbf{1}_{NN} \otimes \nu_{k-1}\nu_{k-1}^\top)(I_{Nn}- K D_C)^\top\\ &+ (K\overline{D}_C) \zeta_k\zeta_k^\top (K\overline{D}_C)^\top
\end{align}
where $\mathbf{1}_{NN}$ is $N$ by $N$ matrix of $1$s. Then,
\begin{align} \nonumber
\Sigma =& (I_{Nn}- K D_C)(\mathbf{1}_{NN} \otimes Q)(I_{Nn}- K D_C)^\top\\ &+ (K\overline{D}_C) R (K\overline{D}_C)^\top
\end{align}
The $2$-norm of $\Sigma$ is upper-bounded by,
\begin{align}
 \nonumber
\|\Sigma\|_2 \leq& \|(I_{Nn}- K D_C)(\mathbf{1}_{NN} \otimes Q)(I_{Nn}- K D_C)^\top\|_2\\ \nonumber
&+ \|(K\overline{D}_C) R (K\overline{D}_C)^\top\|_2 \\
\leq & \|I_{Nn}- K D_C\|_2^2 N \|Q\|_2 + \|K\|_2^2 \|\overline{R}\|_2
\end{align}
where $\overline{R}=\mbox{blockdiag}(C_i^\top R_iC_i)$.
Let $\|I_{Nn}- K D_C\|_2^2=\alpha_1$, $\|K\|_2^2=\alpha_2$, and $\|\overline{R}\|_2=\beta \|R\|_2$. Then, from \eqref{eq_pinfty1} and scaling the error covariance by $N$ (number of sensors),
\begin{eqnarray} \label{eq_pinfty}
\frac{\|P_{\infty}\|_2}{N}  \leq \frac{\alpha_1N\|Q\|_2+\alpha_2\beta \|R\|_2}{N(1-b^2)}=\Phi
\end{eqnarray}

In fact,  equation~\eqref{eq_pinfty} gives an upper-bound on the variance of estimation error at each sensor $\mb{e}_k^i$.
Following the Gaussianity of estimation error, one can claim that with probability more than $68\%$ the estimation error lies within $|\mb{e}_k^i- \mathbb{E}(\mb{e}_k^i )|<\Phi$. It can be proved that in the steady state, the fault-free estimation error is unbiased, i.e. $\mathbb{E}(\mb{e}_k^i )=0$ \cite{khan2014collaborative}. Therefore, one can claim that with probability more than $68\%$ we have  $\mb{r}_k^i=|C_i\mb{e}_k^i+\mb{\zeta}^i_k|<c\Phi+R$
where the constant $c$ is the absolute value of the measurement vector (with the assumption of having one state measurement by every sensor, $c$ is the absolute value of the nonzero entry of $C_i$). Similarly, with probability more than $95\%$, the residual lies below $2c\Phi+2R$, i.e. $\mb{r}_k^i<2c\phi+2R$, and with probability more than $99\%$ we have $\mb{r}_k^i<3c\Phi+3R$, etc. Therefore, in the presence of possible sensor faults, one can detect and isolate sensor faults by comparing thresholds with the residuals based on these probabilities. In other words, considering $T_{68\%}=c\Phi+R$ as threshold, one can claim fault detection with probability $68\%$ whenever the residual goes over this threshold. In this case, the probability of \textit{false alarm} is less than $32\%$. Similarly, stronger thresholds for fault detection and isolation can be defined as $T_{95\%}=2c\Phi+2R$, $T_{99\%}=3c\Phi+3R$, etc.

\section{Sensor Replacement for Observability Recovery}\label{sec_compens}
The measurement of faulty sensors may be compensated in terms of observability recovery and in this direction, the concept of \textit{observational equivalence} is a relevant term \cite{icassp2016,spl17}. The faulty sensor measurement can be replaced by a measurement
of an equivalent state to recover the loss of observability. Note that in this paper, to avoid trivial case, it is assumed that the fault is \textit{inherent} with the measurement of the specific state measured by the faulty sensor. This could be due to, for example, environmental condition of the sensor location resulting to the  fault/anomaly at the sensor.

\begin{figure}
	\centering
	\includegraphics[width=2.5in]{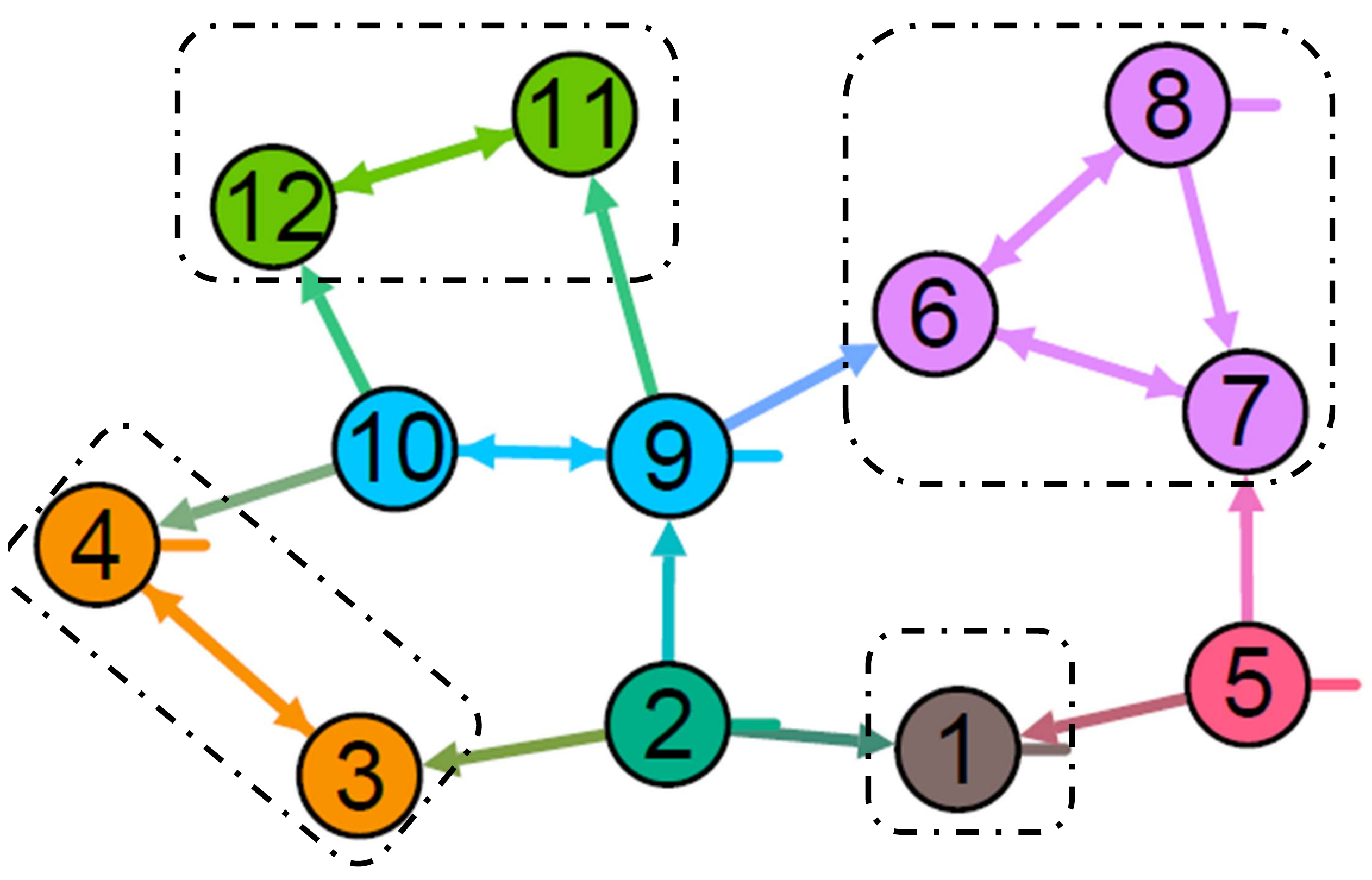}
	\caption{The system digraph associated with the structured system matrix \eqref{eq_A}. The self-cycles are shown by a small line on the right-hand-side of the nodes. Nodes of the same color belong to the same SCC. The SCCs inside the black dashed rectangles are parent SCCs. }
	\label{fig_G}
\end{figure}

Extending the results of \cite{isj_minimal} to dual case of network observability, the set of necessary sensor measurements can be found over digraph representation of the dynamical system.
Note that the \textit{system digraph}, denoted by $\mc{G}$, is defined as the graph associated with the system matrix $A$, or the structured system matrix $\mc{A}$ representing the zero-nonzero pattern of $A$. In $\mc{G}$ every state is represented by a node and every non-zero entry $A_{ij}$ (or $\mc{A}_{ij}$) is represented by a link from node $i$ to node $j$. It is known that many generic properties of the system, including observability and controllability, can be defined over this graph \cite{lin}.  As an example, consider the following structured system matrix,
\begin{align}\label{eq_A}
\mc{A}=\left(
\begin{array}{cccccccccccc}
*& 0& 0& 0& 0 &0 &0& 0& 0 &0& 0& 0\\
*& *& *& 0 &0 &0 &0 &0& * &0 &0 &0\\
0 &0& 0 &*& 0& 0 &0 &0 &0 &0& 0& 0\\
0& 0 &* &* &0& 0 &0 &0 &0 &0 &0 &0\\
* &0 &0 &0 &* &0 &* &0 &0 &0& 0 &0\\
0 &0 &0 &0 &0 &0 &* &* &0 &0 &0 &0\\
0& 0& 0 &0& 0& *& 0& 0& 0& 0& 0& 0\\
0& 0& 0& 0& 0& *& *& *& 0& 0& 0& 0\\
0& 0& 0& 0& 0& *& 0& 0& *& *& *& 0\\
0 &0 &0& * &0& 0& 0& 0& *& 0 &0& *\\
0& 0 &0 &0 &0& 0 &0 &0 &0 &0 &0 &*\\
0& 0& 0& 0& 0& 0& 0& 0& 0& 0& *& 0
\end{array}
\right)
\end{align}
where $*$ represents a non-zero entry and as an example a non-zero entry in $\mc{A}_{51}$ implies a link from node $5$ to node $1$ in $\mc{G}$. The  system digraph $\mc{G}$ associated to \eqref{eq_A} is shown in Fig.~\ref{fig_G}.
In the system digraph, $\mc{G}$,  a Strongly-Connected-Component (SCC) denoted by $\mc{S}$ is defined as a subgraph in which there is a path from every node to every other node. Among the SCCs, define a parent SCC, denoted by $\mc{S}^p$, as the SCC with no outgoing  link to other SCCs. In Fig.~\ref{fig_G}, the SCCs and parent SCCs associated with the structured system matrix \eqref{eq_A} are shown. As it can be seen from this figure, the SCC may only include  a single node.

It can be proved that  the measurement of at least one node in every parent SCC is necessary  for system observability  \cite{isj_minimal}, while all state  nodes in the same parent SCC are observationally equivalent. This implies that the measurements of all nodes in the parent SCC equivalently recover the observability of the system digraph, $\mc{G}$.

In this direction, our proposed logic for  sensor replacement is as follows: \textit{if the faulty sensor measures a state in a parent SCC $\mc{S}^p_i$, one can compensate the loss of observability by adding a new sensor measurement of another state node in $\mc{S}^p_i$. Otherwise, if the faulty sensor measures a state in no parent SCC, the faulty sensor can be removed with no affect on the estimation performance of the other sensors.} For example, in Fig.~\ref{fig_G} if the measurement of state $8$ is faulty and is removed, measurement of either states $6$ or $7$ may recover the loss of observability. Note that if the faulty sensor does not measure a necessary system state (a state in a parent SCC), its removal has no affect on distributed observability, and only the communication network needs to be restructured to satisfy the strong connectivity of the sensor network as mentioned in Section~\ref{sec_est}. For example, in Fig.~\ref{fig_G} a faulty sensor measuring any state in $\{2,5,9,10\}$ can be removed without loss of system observability. It should be noted that if the parent SCC includes a single node (also referred as parent node), there is no replacement for the faulty measurement. This implies that the parent nodes are more vulnerable to attacks and faults in terms of recovering system observability. For example, this is the case for the measurement of state $1$ in Fig.~\ref{fig_G}.  

\section{Simulations} \label{sec_sim}
\subsection{An Illustrative Example}
\begin{figure}
	\centering
	\includegraphics[width=3in]{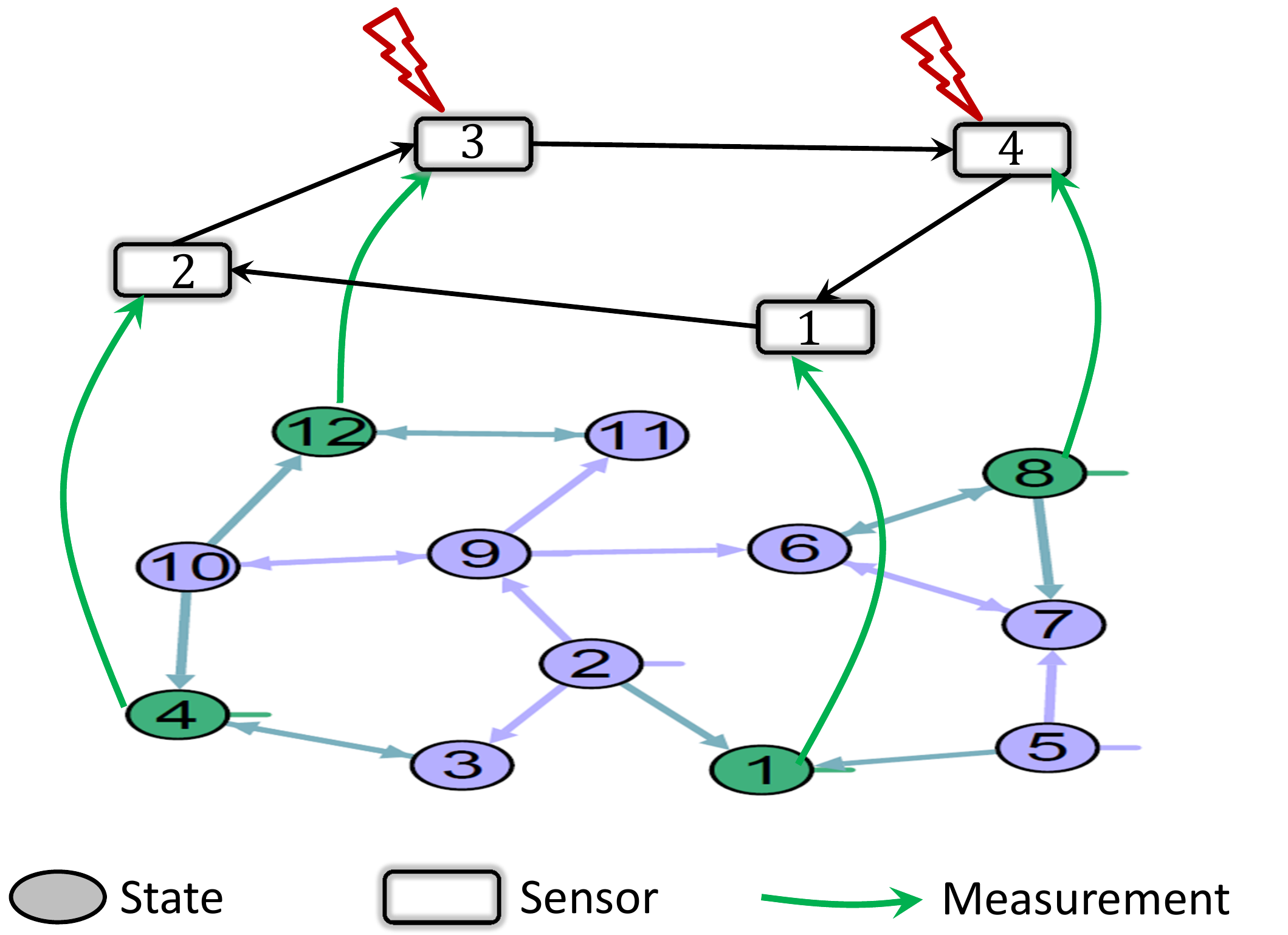}
	\caption{An illustrative example  for networked estimation of dynamical system over a sensor network where the $12$ nodes graph in the bottom is the system digraph associated with a full-rank dynamical system, the green states are measured by sensors, and  the sensor network on top includes a SC communication network of $4$ sensors with faulty sensors $3$ and $4$.}
	\label{fig_system}
\end{figure}

Consider a linear dynamical system with the structure given by the system digraph in Fig.~\ref{fig_system}.
The full-rank system of $12$ states with associated structured matrix \eqref{eq_A} is considered to be tracked by a network of $4$ sensors. The non-zero entries of $\mc{A}$ are chosen such that $\rho(A)=1.2>1$ implying an unstable system dynamics. The system and measurement noise are $\mc{N}(0,0.04)$. The measured states are represented by green nodes. Given the set of measurements as in Fig.~\ref{fig_system}, it can be checked that one state node in every parent SCC $\mc{S}_1^p=\{1\}$, $\mc{S}_2^p=\{3,4\}$, $\mc{S}_3^p=\{6,7,8\}$, $\mc{S}_4^p=\{11,12\}$ is measured and therefore the pair $(A,C)$ is observable. The sensors estimate the system state over time using the networked estimator \eqref{eq_p}-\eqref{eq_m}. The sensors share their \textit{a-priori}  estimates over the SC communication network (or sensor network) shown in Fig.~\ref{fig_system}. This network represents the structure of $W$ matrix while the entries of $W$ are chosen randomly such that the stochasticity of $W$ is satisfied. Having $W$ to be irreducible, for the networked system the conditions for networked observability (or distributed observability) are hold \cite{globalsip14}. Therefore, one can design the proper gain matrix $K$ using the LMI procedure described in Section~\ref{sec_gain}. Applying this block-diagonal $K$ matrix, $\rho(\widehat{A})<1$ and therefore all sensor errors are stable.

\begin{figure}
	\centering
	\includegraphics[width=3in]{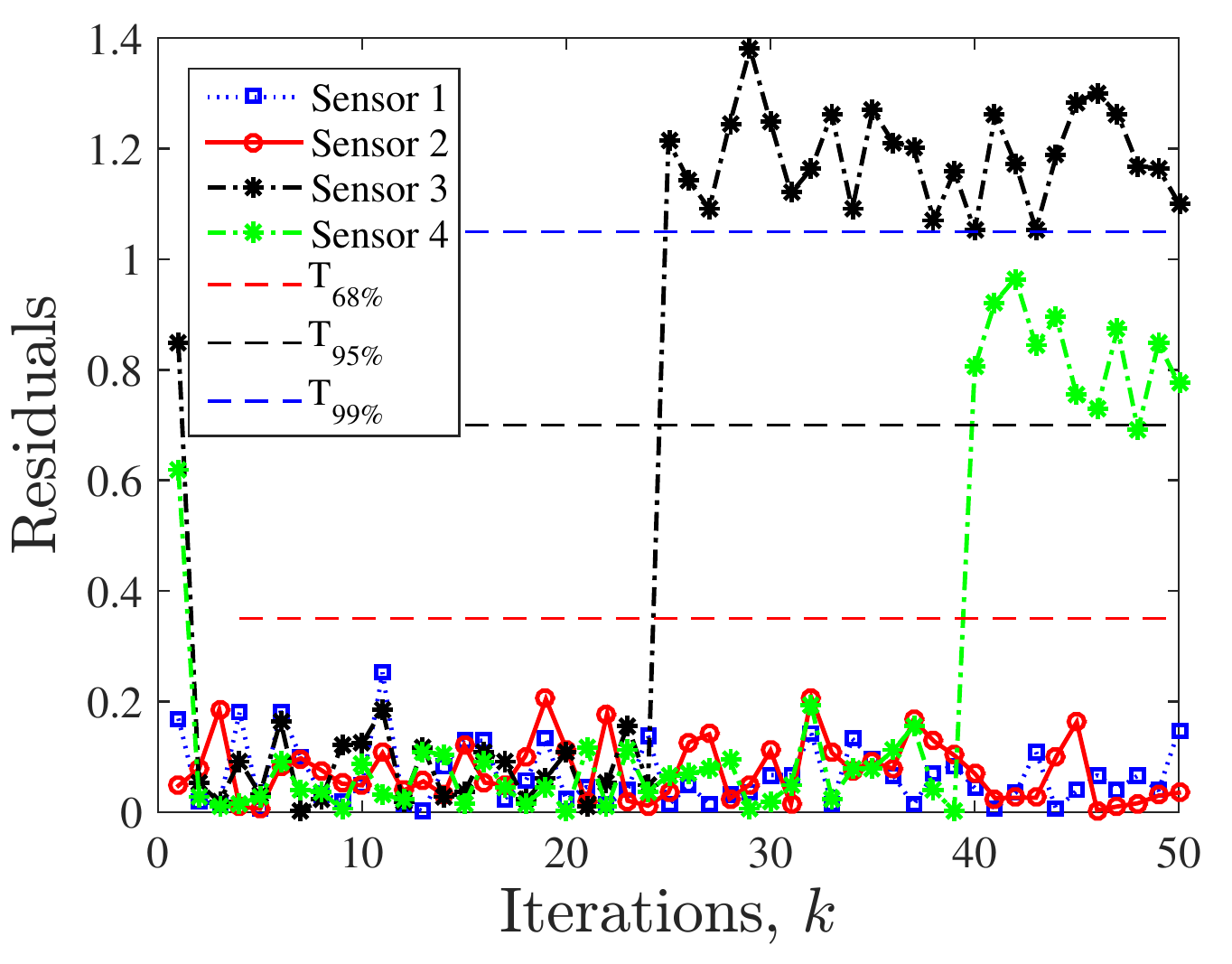}
	\caption{The residuals of $4$ sensors monitoring the system states to detect possible faults at sensors where the residuals corresponding to sensors $3$ and  $4$  exceed  $T_{99\%}$ and $T_{95\%}$ thresholds,  implying that the sensor $3$ and sensor $4$ are faulty, with probability more than  $99\%$ and  $95\%$, respectively.}
	\label{fig_residual}
\end{figure}

Next, it is assumed that the sensor $3$ is faulty after time-step  $25$ and  sensor $4$ is faulty after time-step  $40$, i.e. $\mb{f}^3_{k\geq25}=0.6$, $\mb{f}^4_{k\geq40}=0.4$. The residuals  corresponding to this fault scenario  are shown in Fig.~\ref{fig_residual}.
As shown in this figure, the faulty sensors can be detected and isolated once the residuals corresponding to the faulty sensor exceed the  thresholds. For this simulation, we have $b = \|\widehat{A}\|_2=0.63$,  $\alpha_1 = \|I_{48}- K D_C\|_2^2= 3.83$,  $\alpha_2 = \|K\|_2^2= 3.84$, $|C_3K^3C_3^\top| = 3.01$, and $|C_4K^4C_4^\top| = 3.05$. Further, the non-zero entries of the measurement matrix $C$ are considered to be equal to $1$, and therefore $\beta=1$ and $c=1$. Then, using equation \eqref{eq_pinfty}, it follows that $\Phi=0.31$  and the thresholds are,
\begin{eqnarray}
 T_{68\%} = 0.35,~~
 T_{95\%} = 0.70, ~~
 T_{99\%} = 1.05.
\end{eqnarray}
Based on these thresholds, one can claim the detection of fault at sensor $3$ with probability more than $99\%$ and fault at sensor $4$ with probability more than $95\%$. The probability of false alarms are  less than $1\%$ and less than $5\%$, respectively.

\begin{figure}
 	\centering
 	\includegraphics[width=3in]{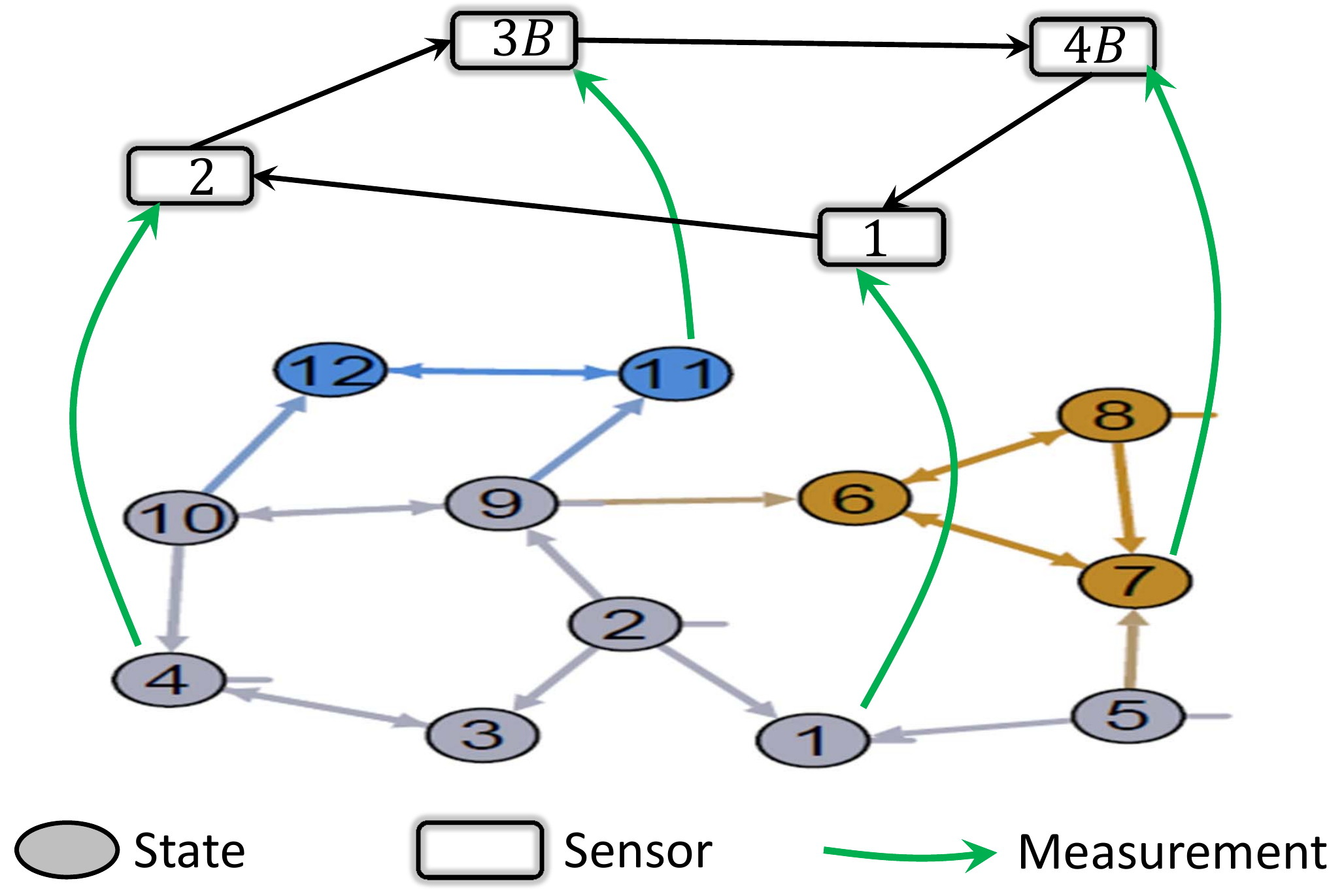}
 	\caption{Observability recovery where  the set of states $\{11,12\}$ and $\{6,7,8\}$ whose measurements are observationally equivalent are shown, respectively,  in blue and brown colors. The faulty sensors $3$ and $4$ in Fig.~\ref{fig_system} are replaced with new sensors $3B$ and $4B$, respectively, measuring the observationally equivalent states $11$ and $7$.}
 	\label{fig_equivalent}
 \end{figure}

\begin{figure}
 	\centering
 	\includegraphics[width=3in]{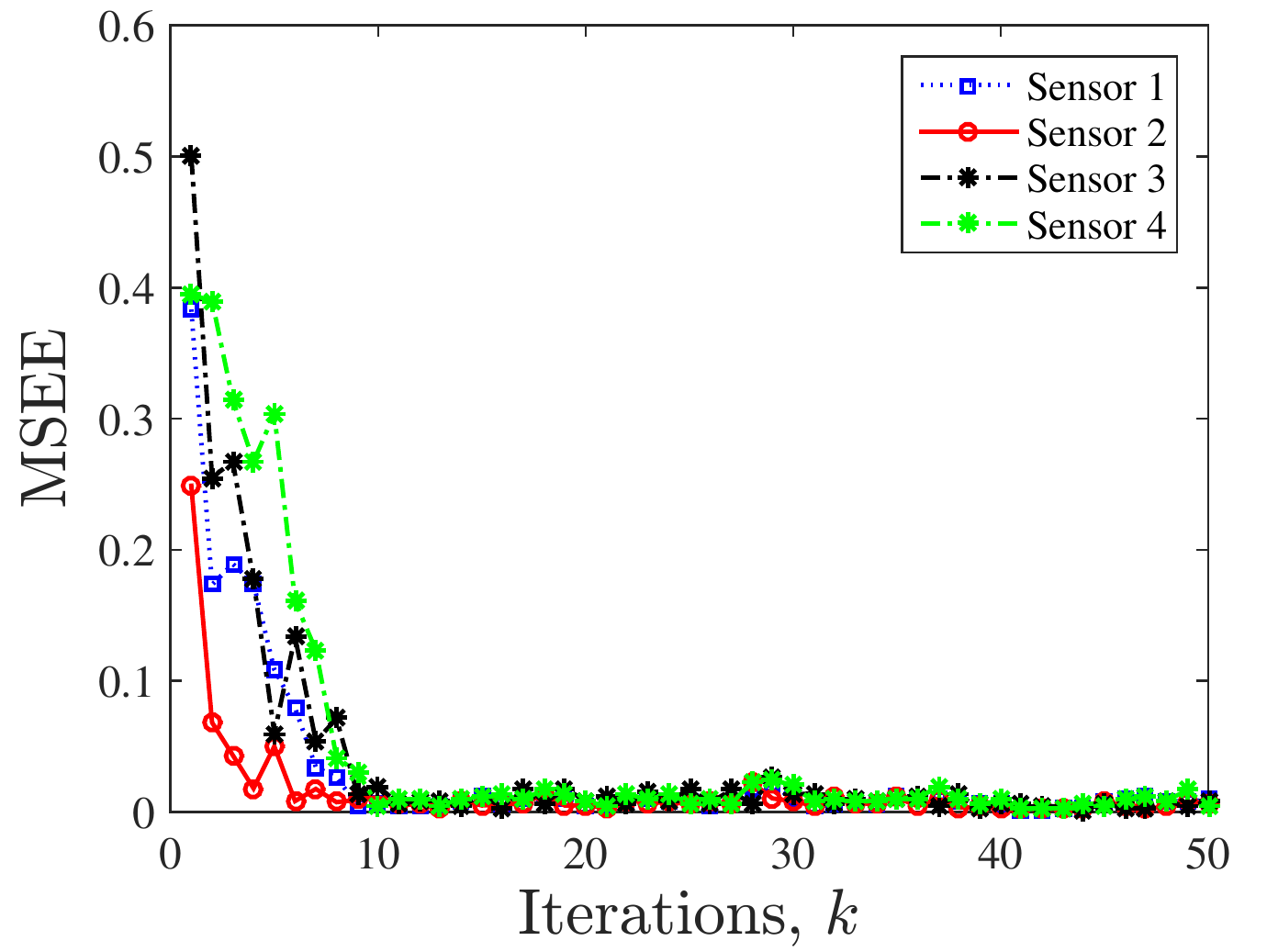}
 	\caption{The MSEE at $4$ sensors given in Fig.~\ref{fig_equivalent} with sensor $3B$ replacing the faulty sensor $3$ and  sensor $4B$ replacing the faulty sensor $4$ in Fig.~\ref{fig_system}.}
 	\label{fig_MSEE}
 \end{figure}

Next, one can compensate for measurement of the faulty sensors by replacing observationally equivalent state measurements from $\mc{S}_3^p=\{6,7,8\}$ and $\mc{S}_4^p=\{11,12\}$.  In Fig.~\ref{fig_equivalent}, the set of states which are observationally equivalent to state $12$ (measured by sensor $3$) and to state $8$ (measured by sensor $4$) are, respectively, shown in blue and brown colors.
Replacing the faulty sensors $3$ and $4$ with sensors $3B$ and $4B$ measuring observationally equivalent states $11$ and $7$, the loss of observability for estimation procedure is recovered. Note that we assume  communication links for sensors $3B$ and $4B$ to be the same as sensor $3$ and $4$. The Mean-Squared-Estimation-Error (MSEE) at all sensors in the compensated framework is shown in Fig.~\ref{fig_MSEE}.  As it is clear the MSEE is bounded steady-state stable at all sensors.

As compared to simulations in \cite{jee2012h,li2009observer,deghat2019detection} where the underlying dynamical system is  \textit{stable}, the example in this section provides a distributed estimation and fault detection of an \textit{unstable} system. As compared to \cite{pajic2015attack,chong2015observability,lee2015secure,kodakkadan2017observer,Riccati-weakcons,marino2017distributed}, in this example the noise follows Gaussian distribution and no bound on the noise is assumed. Moreover, in \cite{teixeira2014distributed,chen2016dynamic,li2016distributed,mitra2018secure}, no system and/or measurement noise is considered. As compared to the distributed estimation in \cite{sauter:09,kar2012distributed,azizi2014networked} which requires a complete all-to-all network of sensors  (i.e. for this example every sensor is directly connected to all other $3$ sensors), our networked estimator only requires an SC sensor network. Among the related literature, we provide a comparison with a recent work in the next subsection.

\subsection{A Comparison Study}
Here, our proposed sensor fault detection, isolation, and distributed estimation protocol is compared with the recent work \cite{he2020secure}. In \cite{he2020secure}, a multi time-scale resilient distributed estimation strategy is proposed along with attack/fault detection. The reason for selecting this work for comparison is that its assumptions and framework is similar to our proposed strategy in the following aspects: (i) it considers the underlying system dynamics to be unstable; (ii) the estimation and attack detection scheme is not centralized but distributed over a sensor network; (iii) it assumes a connected undirected network of sensors, while similarly we assume a SC network; and (iv) an additive term is considered in the presence of noise on the sensors representing possible biasing attacks/faults at sensors similar to our assumption.
Note that in \cite{he2020secure}, each sensor performs $L$ steps of (consensus) averaging over the sensor network between every two steps $k-1$ and $k$ of system dynamics and it is claimed  that the proposed protocol is resilient to sensor faults of certain magnitude, while it is capable of detecting and isolating the attacked sensors in certain conditions. They particularly state that under certain conditions the attacked sensor can be detected while some other attacks cannot be detected (remain stealthy), and therefore two sets of \textit{detectable attacking set} (DAS) and \textit{undetectable
(or stealthy) attacking set} (UAS) are defined. The proposed strategy in \cite{he2020secure} is simulated over the same dynamic system and sensor network in Fig.~\ref{fig_system} and all the conditions including initial state values, noise values, etc are considered similar to the previous subsection, while the sensor network is reconsidered as an undirected cycle $1 \leftrightarrow 2 \leftrightarrow 3 \leftrightarrow 4 \leftrightarrow 1 $. The parameters for the distributed estimation protocol are  as in Table~\ref{tab_1}.
\begin{table} [hbpt!]
	\centering
	\caption{Prameter values for resilient distributed estimation and attack detection protocol in \cite{he2020secure}.}
	\begin{tabular}{|c|c||c|c||c|c|}
		\hline
		$L$& $20$ & $\alpha$ & $0.2$ &$\beta$ & $0.5$ \\
		\hline
		$\|A\|$ &  $1.37$ & $\gamma$ & $0.21$  & $N$& $4$ \\
		\hline
		$s$ & $2$  &$b_w$ & $0.1$ &$b_v$ & $0.1$ \\
		\hline
		$\lambda_0$ & $1.9$  &$\eta_0$ & $0.1$ &$\rho_{t_0}$ & $0.1$ \\		
		\hline
		\hline
	\end{tabular}
	\label{tab_1}
\end{table}

\begin{figure}
 	\centering
 	\includegraphics[width=3in]{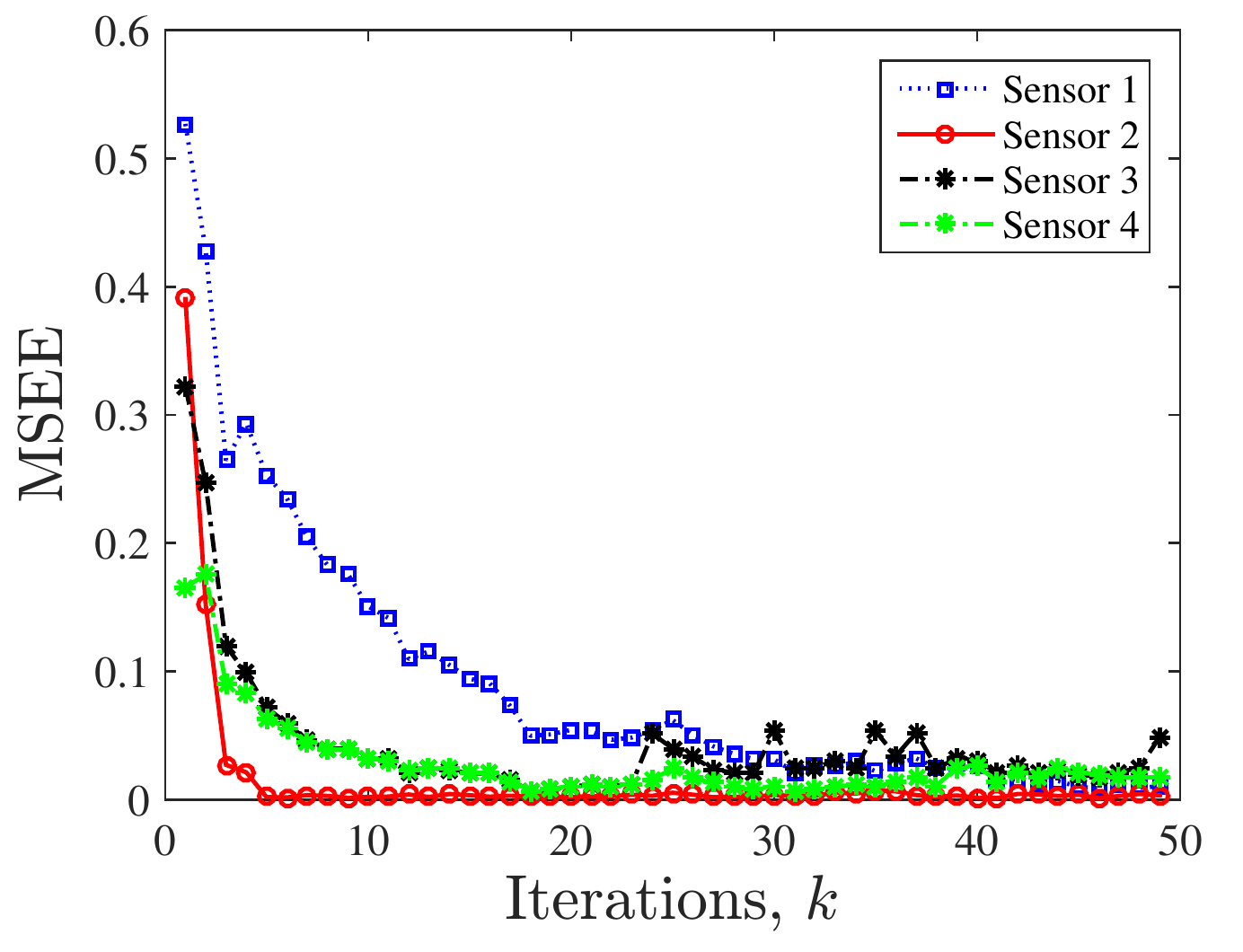}
 	\caption{The MSEE at $4$ sensors state estimation under the proposed  protocol in \cite{he2020secure}.}
 	\label{fig_msee2}
 \end{figure}

\begin{figure}
 	\centering
 	\includegraphics[width=3in]{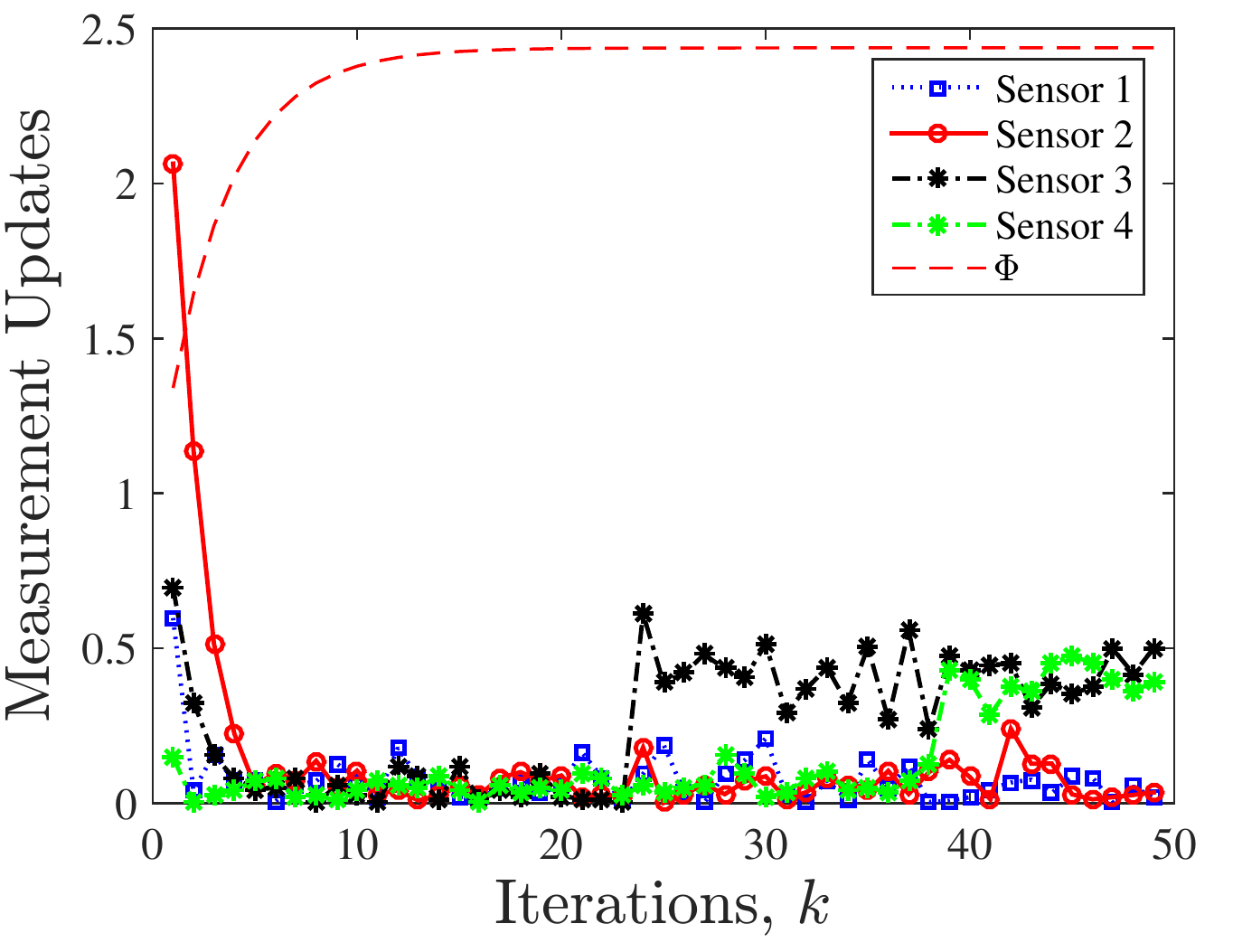}
 	\caption{The threshold $\Phi_k$ and the measurement updates  at $4$ sensors estimating the system  in  Fig.~\ref{fig_system} under the proposed attack detection scheme in \cite{he2020secure}. As claimed in \cite{he2020secure} the attack/fault is detected if the measurement updates exceed the $\Phi_k$. In this case, both attacks/faults at sensor $3$ and $4$  remain undetected (stealthy). }
 	\label{fig_detect2}
\end{figure}

Under these parameters the MSEE of  $4$ sensors are shown in Fig.~\ref{fig_msee2}. Note that in this simulation, every sensor performs $L=20$ steps of consensus on estimations of neighboring sensors as compared to $1$ step consensus in protocol \eqref{eq_p}-\eqref{eq_m}. This requires a processing setup $20$ times faster than our proposed networked estimation setup.
The sensor attack detection logic in \cite{he2020secure} is as follows: if the measurement update $\mb{y}^i_k-C_iA\widehat{\mb{x}}^i_{k-1}$ is larger than a pre-defined threshold $\Phi_k$, then the attack at sensor $i$ is detected, otherwise the sensor is either attack-free or the attack remains undetected (stealthy). Running the proposed attack detection in \cite{he2020secure} based on the parameters in Table~\ref{tab_1}, the measurement updates and the threshold $\Phi_k$ are shown in Fig.~\ref{fig_detect2}. As it is clear from this figure, both attacks/faults at sensors $3$ and $4$ are not large enough to be detected and both remain stealthy. However, as shown in Fig. \ref{fig_residual}, our proposed approach can detect and isolate both faults.

\section{Concluding Remarks}\label{sec_con}
In this paper, a fault detection and isolation scenario for a networked estimator is presented by  defining the sensor residuals and probability-based thresholds to detect and isolate sensor faults. The thresholds are defined based on the upper-bound on the norm of the estimation error covariance. It should be emphasized that the  FDI and distributed estimation strategy in this paper is of polynomial order computational complexity. In fact,  the protocol \eqref{eq_p}-\eqref{eq_m} as a variant of the protocol in \cite{isj_cyber} is of P-order complexity and according to \cite{ye1993fully,nesterov1994interior}, Algorithm~\ref{algorithm} and similar cone-complementarity algorithms are of P-order complexity, implying that the design of block-diagonal gain matrix in Section~\ref{sec_gain} is of P-order complexity. For fault detection and mitigation strategy, the decomposition of the system digraph into SCCs, determining their partial order, and finding parent SCCs are based on the Depth-First-Search (DFS) algorithm \cite{algorithm} with complexity $\mc{O}(n^2)$. The computational complexity of the threshold design in \eqref{eq_pinfty} based on the 2-norm of the covariance is $\mc{O}(n^3)$. Therefore, the overall strategy in this paper is of P-order complexity and hence scalable  to large-scale applications. Note that although the simulation in Section~\ref{sec_sim} is given for a small-scale example system, the P-order complexity guarantees the scalability of the proposed scheme to large-scale systems.

It is worth noting that the fault detection and isolation can be improved by defining tighter upper-bounds on the residuals. This can be done by designing the gain matrix $K$  to reduce  $\|I_{Nn}-KD_C\|_2^2$, and $\|K\|_2^2$.  The optimal LMI approach to optimize the gain $K$ to satisfy such conditions is the direction of our future research.
A promising topic of future interest is optimal design of the communication network (the structure of the matrix $W$) and cost-optimal  selection of measurements for observability recovery to reduce the estimation costs at sensor network. This may include reducing both the sensor embedding costs and communication costs in sensor networks. As another direction of future research we are currently working to extend this fault detection  and isolation as well as fault compensation scenario to general rank-deficient systems. In this direction, the networked estimation protocol needs to be updated to consider measurement sharing over the sensor network.

\section*{Acknowledgement}
We would like to thank Prof. Usman A. Khan from Tufts University for his helpful comments and suggestions on this paper. We also thank the authors of reference \cite{he2020secure} for providing their paper and related files.

\bibliographystyle{IEEEtran}
\bibliography{bibliography}

\end{document}